\newcommand{\yr}{{\rm yr}}
\newcommand{\msolar}{{{\rm M}_\odot}}
\newcommand{\lsolar}{{{\rm L}_\odot}}
\newcommand{\kelvin}{~^0{\rm K}}
\newcommand{\be}{\begin{equation}}
\newcommand{\ee}{\end{equation}}
\newcommand{\beqar}{\begin{eqnarray}}
\newcommand{\eeqar}{\end{eqnarray}}
\newcommand{\bear}{\begin{array}}
\newcommand{\ear}{\end{array}}
\documentstyle{article}
\begin{document}
\baselineskip=15pt

\begin{center}
COOLING FLOW MODELS OF THE X--RAY EMISSION AND TEMPERATURE PROFILES FOR A 
SAMPLE OF ELLIPTICAL GALAXIES 
\end{center}

\vspace*{3.0truecm}

\begin{center}
G. Bertin and T. Toniazzo 
\end{center}

\vspace*{2.0truecm}

Scuola Normale Superiore, Piazza dei Cavalieri 7, I 56126, Pisa, Italy\\

\vspace*{3.0truecm}

\vspace{2,5truecm}
\noindent
ABSTRACT\\

\noindent

A simple spherically-symmetric, steady-state, cooling-flow 
description with gas loss (following Sarazin \& Ashe  1989), within galaxy
models
constrained by radially extended stellar dynamical data, is shown to provide
generally reasonable fits to the existing data on X-ray emission profiles and
temperatures for a set of bright elliptical galaxies in Virgo and Fornax. Three
free parameters are needed to specify the model: the external mass flux, the
external pressure, and a dimensionless factor, which
regulates the mass deposition rate along the flow. Three different assumptions
on the supernova rate have been considered. A moderate value for the
supernova rate in elliptical galaxies is found to be preferred. Confining
pressures of $p_{ext}\sim4\div15\times10^3\kelvin$ cm$^{-3}$ and  significant
accretion rates of external material, up to $4\msolar/\yr$, are suggested by 
our models. A possible correlation between $L_X/L_B$ and the iron abundance 
in the gas inside ellipticals is pointed out.

\vspace{2.0cm}
\noindent
Subject headings:

Galaxies: Cooling Flows -- Galaxies: Elliptical and Lenticular, cD -- Galaxies:
Intergalactic Medium -- Galaxies: ISM -- Galaxies: Structure -- X-Rays:
Galaxies\\

\vspace{2,5truecm}
\noindent
1. INTRODUCTION\\

\noindent

In their simplest form, cooling flow models for the description of the X--ray
emission from individual  galaxies (see Sarazin and White 1987, Vedder, 
Trester, and Canizares 1988) have a history of mixed success (see Sarazin 
1990 and references therein). 

On the one hand, a steady state description 
of the dynamics of the hot interstellar medium may be too naive, in view of 
the expected time dependence of the basic energy balance during galaxy 
evolution (see Loewenstein and Mathews 1987, David, Forman, and Jones 1990, 
Ciotti et al. 1991, Binney and Tabor 1995). In this respect, as with several 
examples in other 
physical contexts, one argues that the time scale of the overall evolution 
of the system that is considered is sufficiently  long that a simple steady 
state analysis can correctly capture the properties of at least some of the 
stages of the time evolving hot gas, and, in particular, that the steady  
state cooling "inflow" description gives a reasonable representation of 
the current state of the brightest X-ray ellipticals. Attempts have been 
made to check in detail what are the conditions for the applicability of the 
steady state description through examination of the results of 
hydrodynamical codes (Murray and Balbus 1992). 

On the other hand, it was 
immediately realized that the simplest cooling flow models all suffered 
from a basic inconsistency with the observations in that, given the piling 
up of hot gas, the theoretically predicted emission profiles are too bright 
close to the galaxy center (see Sarazin and White 1988). Various ways have 
been proposed to correct the cooling flow model in order to resolve the 
problem of the excessive steepness of the surface brightness profile, by
recognizing that part of the hot  gas is bound to decouple away from the cooling
flow (Thomas 1986, Thomas et  al. 1986, Sarazin and Ashe 1989). To some extent,
the  filamentary structure 
of the gas noted earlier especially around cD's (e.g., see Lynds 1970) and 
dramatically demonstrated on the cluster scale by recent {\it Rosat} 
observations (Sarazin, O'Connell, and McNamara 1992a,b) adds further 
encouragement in this direction. One might even conclude that some of these 
observations undermine the very foundation of the simple spherical, steady 
state, one--phase cooling flow as a model for the hot gas.

The impact of the constraints posed by stellar dynamical data 
on the modeling of the X--ray emitting gas in elliptical galaxies was 
examined in a previous paper (Bertin, Pignatelli, and Saglia 1993; 
hereafter BPS). This paper focused 
on the problem of dark matter and on the galaxy NGC 4472, for which 
radially extended optical spectroscopic profiles are available (Saglia et 
al. 1993). Such a study gave one surprising result and 
confirmed one well known fact. The surprise was to find that the 
predicted temperature and surface brightness profiles for the hot gas for models
with or without  significant amounts of dark matter, if constrained by the same
set of  stellar dynamical data out to $\approx R_e$ (the {\it effective}
optical 
radius), are very similar to each other even out to $\approx 7 R_e$; in 
particular, the temperature profiles are found to be much more sensitive to 
the assumed value of the intracluster pressure $p_{ext}$ than to the amount 
of dark matter that may be argued to be present. This surprising result 
urges caution in the use of X--ray data as diagnostics for the presence of 
dark halos (see Loewenstein 1992 and Serlemitsos et al. 1993). The well 
known fact confirmed by the analysis of BPS is the central overbrightness 
problem for the simplest form of cooling flow models of the hot gas: the 
case of NGC 4472 shows that the use of accurate models for the 
distribution of the stellar component can only lead to a slight improvement 
of the predicted X--ray emission profile, which remains too steep and 
definitely fails at $R \leq 0.2 R_e$. 

In this paper we continue the study started by BPS, by focusing on the 
modeling of the X--ray emitting gas for galaxies for which accurate stellar 
dynamical models are available. In doing so, we may have confidence in the 
adopted profiles for the functions $\rho_{\star}(r)$ and $\sigma_{\star}(r)$, 
that characterize the stellar density and velocity dispersion profiles, and 
for the gravitational potential $\Phi(r)$. In view of the study of BPS, 
here we omit further discussions on the role of dark matter and simply 
refer to the available best fit stellar dynamical models (which include, in 
general, a dark halo). The main goal of this paper is then to test the 
adequacy of simple steady state, spherical cooling flow models for the hot 
gas by studying in detail the role of the decoupling of cold gas from the 
flow, following the prescription of Sarazin and Ashe (1989), and  the role 
of the intracluster medium. The analysis thus basically relies on the 
variation of three parameters ($q$, $p_{ext}$, and $\dot{m}_{{ext}}$; see 
\S 3 below). Qualitatively, the effects of these variations are well known 
(see papers cited above). Here we would like to make a full detailed 
quantitative test in order to ascertain whether a simple uniform choice for 
$q$ and physically plausible properties of the intracluster medium can lead 
to good fits to the X--ray emission profiles and to the temperature 
profiles (if available) for several objects.

As noted earlier (e.g., see discussion in \S 2 of BPS), the presence of a 
finite pressure 
at the outer boundary is associated with the existence of 
an inflow of gas from the cluster. 
Furthermore,  
for some objects (like NGC1399 and NGC4636 in our sample, see \S 4), 
a significant rate of accretion from the cluster, exceeding the value 
of 1$\msolar/\yr$, is found to be required in the present frame of 
steady--state cooling flows. Therefore, in this paper, as will be described in 
the following \S 3, 
we solve the relevant equations treating the outer boundary
differently from Vedder et al. (1988)  and BPS, in that 
we consider the possibility of an inflow 
of extragalactic matter as a separate parameter ($\dot{m}_{{ext}}$), 
in addition to the pressure at the external boundary $p_{ext}$, 
to be adjusted in the fit process. For galaxies with a low value of 
$L_X/L_B$ for which no significant accretion is required, in order to avoid
the formal singular behavior of the solution which would be associated with the
so--called stagnation radius (introduced as the radius  where the radial
velocity of the flow vanishes), we set  $\dot m_{ext}=0.1 \msolar/\yr$.
When the 
accretion rate is so low, the resulting profiles are  hardly affected by 
the precise value of $\dot{m}_{{ext}}$.

The physics  of the hot gas is intrinsically complex. In this paper we 
focus on three factors of the problem, on the basis of 
their physical interest. The first factor is the decoupling of cold gas from the
hot inflow.  The mass decoupling term introduced by Sarazin and Ashe (1989) 
provides a phenomenological description (with one 
free parameter, called $q$) of the consequences of 
small-scale fluid instabilities which are expected to take place in 
the cooling gas 
(although different views can be taken on the issue; 
see Fabian and Nulsen 1977, Mathews and Bregman  1978, Balbus 
1988, Loewenstein 1989, 1990). 
Therefore, 
it would be desirable that the fitting value of $q$ be of order unity, with 
small variations from one galaxy to another, 
and that the model profiles 
turn out to be not too sensitive to $q$ in the vicinity 
of the best fit value. 
The second factor is the role of the external pressure, which can be 
constrained, to some extent, by direct observations of the intracluster 
plasma.
By considering  galaxies in the same 
cluster, i.e. embedded in the same intracluster medium, one can better 
appreciate whether the adopted values of 
$p_{ext}$ and $\dot{m}_{ext}$ are physically plausible. Ideally, 
besides studying the case of 
different clusters (as is done below), it would be very interesting to 
model truly isolated X--ray bright ellipticals; unfortunately, such a case 
is not easy to find. The third physical factor of the problem addressed in 
the present analysis is related to the determination of the 
supernova Ia rate in ellipticals. The cooling flow models found in this 
paper appear to give better fits to the data if a value is 
used consistent with the recent estimates of Cappellaro et al. (1993).

Even if the set of objects modeled in this paper is very small,
the results of 
the fits performed here (\S 4) are of considerable interest. It is 
found that reasonable fits are obtained with a reasonable choice of the 
parameters involved, {\it even if the models used are highly idealized}. 
Furthermore, if we consider that 
the data on X--ray  temperature "profiles" used here are 
very crude (with the exception of the {\it Rosat} data for NGC 4636, taken 
from Trinchieri et al. 1994),  and 
that the reduction of X-ray 
spectroscopic data and their interpretation usually leave several options 
open (that is, the choice of the parameters to be used in the fit 
such as chemical composition, HI 
absorption, and number of emitting components involved; see Trinchieri et al. 
1994 and Pellegrini \& Fabbiano 1994), 
we may take the theoretical models produced in this paper as 
"predictions" against the better temperature profiles that are currently 
being gathered with the help of new telescopes and instruments. 

\vspace{2,5truecm}
\noindent
2. THE SAMPLE.\\

\noindent
The galaxies that define our sample have been selected on the basis of the 
following criteria: (i) High X-ray luminosity, (ii) Availability of good 
quality X-ray data, (iii) Availability of good quality stellar dynamical data 
and of detailed global stellar dynamical models. 
Criterion (i) is related to our plan to  model the X-ray 
emission and temperature profiles in terms of cooling flows; this 
is justified only for objects with high $L_X/L_B$ 
(see, e.g., Ciotti et al. 1991). In relation to point (ii), 
we have referred to the X-ray contour maps and 
circularized surface brightness profiles given in the {\it Einstein} 
X-ray catalogue of Fabbiano, Kim \& Trinchieri (1992), in order 
to pick out galaxies with high signal, extended X-ray halos. Requirement 
(iii) allows us to perform a study of the coronal 
gas within an accurately determined potential well and 
a detailed 
quantitative framework for the properties of the stellar component. 
For this latter purpose we have made use of the results of the 
survey of Saglia, Bertin, and Stiavelli (1992; hereafter SBS). 

The sample of selected objects thus contains 5 galaxies: 
NGC 1399, NGC 1404, NGC 4374, NGC 4472, and NGC 4636. Their
optical and X-ray properties are listed in Table 1. All the galaxies 
selected are fairly round in their optical appearance (this was indeed one of 
the main criteria for the selection of the sample of SBS), which makes the 
use of spherical models at least a reasonable starting point. The last four 
columns of Table 1 describe some X-ray properties of the galaxies, taken 
from the literature. The radius $R_X$ denotes the radius of the X-ray 
emitting region considered by each source, and $L_X$ is the total X-ray 
flux from within $R_X$. The seventh column is the best-fit temperature of a 
single-component Raymond thermal spectrum fitted to the region considered, 
and can thus be taken as an estimate of an average temperature of the 
gas within $R_X$. Finally, it should be noted 
that the reported total X-ray fluxes in the specified bands are highly 
dependent on the assumed hydrogen column density along the line-of-sight.
Fabbiano et al. (1992) estimate the count/energy conversion factor by assuming
a gas temperature of 1 keV and galactic absorption. 

The available X-ray 
emission profiles of NGC 1399, NGC 4472, and NGC 4636 considered in this paper 
are quite accurate. 
We refer to IPC data for NGC 1399, combined IPC/HRI data for NGC 4472 
and {\it Rosat} PSPC data for NGC 4636. Temperature 
determinations are taken from Serlemitsos et al. (1993; {\it BBXRT} 
data for NGC 1399 and NGC 4472), Forman, Jones, and Tucker (1985; {\it Einstein}
IPC data  for NGC 4472) and Trinchieri et al. (1994; {\it Rosat} PSPC data for
NGC 4636). 

NGC 1404 is embedded in the very extended bright halo of 
NGC 1399. Thus the X-ray total luminosity and the emission profile for 
NGC 1404 may suffer from subtraction problems; furthermore, the temperature 
determination is rather uncertain, given the low energy resolution of the 
IPC instrument (see Kim, Fabbiano \& Trinchieri 1992). Similar problems 
might affect the X-ray 
properties reported for NGC4374 (which is near NGC 4406), although HRI data 
are also available for this object. Some of these uncertainties might 
be resolved with the help of {\it Rosat} data, when analyses of these data 
have been published. 

NGC 1399 and NGC 1404 are at the center of the Fornax cluster, while the 
remaining three objects are members of Virgo. 
The choice of the parameter $p_{ext}$ for those galaxies sharing the same 
environment should be made consistent with the presence of a common 
intracluster medium and with the available observations. 
For example, {\it Ginga} (Awaki et al. 
1991, Ikebe et al. 1992) and {\it Rosat} data (B\"{o}hringer et al.
1994) suggest a temperature $T_{ext}$ in excess of $\sim2~$keV for the 
intracluster medium in  Fornax and Virgo.  

The basic properties of the best fit stellar dynamical models adopted in 
the following study are summarized in Table 2. 
As shown by BPS, even if the stellar dynamical data are far from identifying 
a unique global model for the optical galaxy, it is sufficient to focus our 
attention on the best fit model for the mass distribution provided by stellar 
dynamics, since other models compatible with the same set of stellar 
dynamical data are not expected to produce significant changes in the 
cooling flow modeling. 

In general, the models contain 
significant amounts of dark matter, as shown by Figure 1, which illustrates 
the profiles of the cumulative mass-to-light 
ratio for each galaxy. Table 2 lists the mass-to-light ratio
$M_{\star}/L_B$ for 
the luminous component and the total mass-to-light ratio $M/L_B$. The total 
luminosities given in the Table are taken from SBS. 
There is a wide range in the lengthscales $r_L$ (the half-mass radius 
for the luminous component) and $r_D$ (same, but for the dark matter), 
ranging from $r_L=r_D=13~$kpc for NGC 1404 to $r_L=28~$kpc, 
$r_D=165~$kpc for NGC 4636 (columns 4 and 7 in the Table). 
Significant differences are also noted in the value of 
the (one-dimensional) central stellar velocity dispersion $\sigma_\star(0)$ 
(column 8). The galaxy $M/L_B$ ratio and the scale length of the {\it total}
mass  distribution influence the steepness of the X-ray 
emission profile, while the central gas temperature depends upon the depth 
of the potential well, i.e., on the value of $\sigma_\star(0)$. 
So we would expect that objects with a concentrated mass distribution, 
like NGC 1404 and NGC 4374, have a less extended X-ray halo 
than `giants' like NGC 1399 or NGC 4636. Furthermore, as a result of its 
relatively low 
velocity dispersion, NGC 1404, which is embedded in the X-ray halo of 
NGC 1399, should appear as a {\it cold} X-ray source relative to its 
environment; the opposite should occur for NGC 4374. 

Finally, consider the values of $L_X/L_B$ (see last column of Table 1). 
Even for this selected sample of bright E galaxies 
there is a very large spread (a factor 18 between NGC 1399 and NGC 4374). 
Within the cooling flow description, such a wide spread may just reflect a 
different accretion rate from the intracluster medium; in particular, 
NGC 1399 is expected to accrete large amounts of hot gas from its 
surroundings. This point is discussed further in the next section.

\vspace{2,5truecm}
\noindent
3. THE METHOD.\\

\noindent
We describe the hot coronal gas by means of a standard set of equations 
under the assumption of a steady-state, spherically symmetric inflow in the 
presence of mass and energy source terms (related to mass and energy 
injection in the form of hot gas originating from stars and supernovae), of
radiative losses  (responsible for the observed X-ray emission), and  
of a mass sink term (describing the decoupling of cold gas from the hot  
phase; Sarazin \& Ashe 1989):

\be
   \frac{1}{r^2}\frac{{\rm d~}}{{\rm d}r}(r^2 \rho u)=
   \alpha \rho_\star- q\frac{\rho}{\tau_{dec}}
\label{mass}\ee
\be
   \rho u\frac{{\rm d}u}{{\rm d}r}+\frac{{\rm d}p}{{\rm d}r}=
   -\rho\frac{{\rm d}\Phi}{{\rm d}r}-\alpha \rho_\star u 
\label{euler}\ee
\beqar
     \lefteqn{
     \frac{1}{r^2}\frac{{\rm d~}}{{\rm d}r}\left[r^2 \rho u\left(
     \frac{1}{2} u^2 + \frac{5}{2}\frac{p}{\rho} + \Phi \right)\right]=}
   \nonumber \\
      & & = - \rho^2\Lambda(T)
      + \alpha\rho_\star\left(\epsilon_{inj}+\Phi\right)
      - q\frac{\rho}{\tau_{dec}}\left(\frac{1}{2}u^2
      + \frac{5}{2}\frac{p}{\rho}+\Phi\right).
\label{enthalpy}\eeqar

These equations reduce
to the same set used by BPS when $q=0$. Three unknown functions $\rho$, $p$, and
$u$ (mass density,  pressure, and radial velocity) describing the hot, X-ray
emitting gas are  to be derived by solving the above equations under the
appropriate boundary  conditions (see below). The gas temperature $T$ is
obtained from the  equation of state $p=\rho kT/\mu m_p$. Note that we are
ignoring the  contribution of the possible presence of magnetic fields and of
various  transport processes within the gas. In the following, we will refer 
to the solutions of this set of equations as ``$q$-models''.

\noindent
(i)     {\it Input parameters for the hot gas}

The chemical composition of the gas is 
assumed to be ``cosmic'' (Allen 1973), with a mean mass per particle
$\mu=0.63$ in units of the proton mass $m_p$. The quantity   
$\alpha=4.75\times10^{-19}(M_{\star}/L_B)^{-1}$ sec$^{-1}$  (with the
mass--to--light $M_{\star}/L_B$ ratio expressed in solar units) is the rate  of
mass loss of a 15 Gyr old stellar population that is taken to be  ``quiescent''
(i.e., with no significant star formation after  the initial stage), as given by
Renzini (1988). The  specific energy of the injected material is derived from 
$\displaystyle \epsilon_{inj}=\frac{3}{2}(\sigma_\star^2+
\vartheta_{SN}\varepsilon_{SN}\frac{ k\hat T_{SN}}{\mu m_p})$. 
Here $\sigma_\star$ is the 
one-dimensional stellar velocity dispersion (averaged over the 
three directions), taken from the stellar dynamical model, 
$\vartheta_{SN}$ is the SNIa rate relative to Tammann's (1982) value 
(0.22 SNIa per century per $10^{10} \lsolar_{B}$), $\varepsilon_{SN}$ is 
the mechanical energy released in a SNIa explosion normalized to 
$10^{51}$~erg, and $\hat T_{SN}=2.8\times10^7\kelvin$ is an effective 
temperature associated with the hot supernova ejecta mixing into the gas. 
Our reference case will be characterized by 
$\vartheta_{SN}=1/4$ (see Cappellaro et al. 1993) 
and $\varepsilon_{SN}=1$, but we will explore also the regime of very low
supernova rate $\vartheta_{SN}=1/16$, for which the heating is
typically dominated by the stellar contribution, and the regime of high
supernova
rate, $\vartheta_{SN}=1.1$, where the heating is generally supernova dominated.
  
The cooling function $\Lambda(T)$ is obtained by numerical interpolation  of the
results of Raymond, Cox \& Smith (1976), and is suitable for an  optically thin
thermal plasma with cosmic abundances at temperatures 
$10^5\kelvin<T<10^8\kelvin$.

\noindent
(ii)    {\it The mass decoupling term}

Following the physical prescription of Sarazin \& Ashe (1989), we describe 
the decoupling of cold material from the hot gas by referring to the 
timescale
\be
   \tau_{dec}=\frac{5}{2}\frac{p}{\rho^2\Lambda(T)}
   \left(2-\frac{{\rm d~ln}\Lambda}{{\rm d~ln}T}\right)^{-1}
   =\tau_{cool}\left(2-\frac{{\rm d~ln}\Lambda}{{\rm d~ln}T}\right)^{-1}
\label{thintime}
\ee 
which is the linear growth rate of the thermal instability for comoving, 
isobaric density perturbations to the smooth (`homogeneous') flow 
(Mathews \& Bregman 1978). 
In contrast to the injected mass from the stars, the decoupled cold material 
is assumed not to exchange momentum or entropy with the hot gas. The 
dimensionless parameter $q$ appearing in the continuity equation is 
considered to be as a free (phenomenological) constant 
parameter of the model, to be determined by the fit process. 
Note that the decoupling time given above differs from that of most 
models by Sarazin \& Ashe (1989) (hereafter ``$q_S$--models'') 
by a factor
\be 
c(T) = \left(2-\frac{{\rm d~ln}\Lambda}{{\rm d~ln}T}\right),
\label{cfactor}
\ee  
which ranges from 2.0 to 3.9 when $T$ ranges from $0.5$ to 
$1.5$ keV. Because of this temperature dependence, the $q_S$-models behave 
as q-models with a {\it differential} $q=q(r)$. In practice, $q_S$--models 
with $q_S=3q$ and with the values of $p_{ext}$ and 
$\dot m_{ext}$  suggested by 
our best-fit models for NGC 1399, NGC 4472 and NGC 4636, are found to differ
only  slightly from the corresponding $q$-models.

\noindent
(iii)   {\it The stellar component and the gravitational potential}

The three functions $\rho_\star(r)$, $\sigma_\star(r)$, and $\Phi(r)$ are the 
{\it stellar} mass density, the one-dimensional stellar velocity dispersion 
(averaged over the three directions), and the 
gravitational potential of the galaxy (with the contribution of stars and 
dark matter, but, for simplicity, {\it without} the contribution of the gas)
derived from the best-fit stellar dynamical models described in \S 2.

\noindent
(iv)    {\it Boundary conditions and integration scheme}

Equation (1) is integrated to give for the inward mass--flux $\dot m$:
\be
   \dot m = - 4\pi r^2 \rho u = \alpha \left[M_\star (r_{{ext}}) - M_\star (r)
   \right] - q \left[{\it D}(r_{{ext}})-{\it D}(r)\right]  + \dot m_{{ext}},
\label{mdot}\ee
where $M_\star(r)$ is the galactic {\it stellar} mass inside a sphere of 
radius $r$, and \\ 
${\displaystyle D(r)=\int_0^r \frac{\rho}{\tau_{dec}} 4\pi r^2 {\rm d}r}$. 
The sign of $\dot m$ is chosen so that it is positive for an inflow. The 
quantity $\dot m_{ext}$ is the contribution to the flux due to accretion of 
external gas. Its value sets one boundary condition at $r_{ext}$. 
For the models shown in \S 4 , $r_{ext}$ is chosen to be equal to $R_X$ (see 
Table 1). Note that in some objects (like NGC 4374) this radius 
may fall within the optical galaxy. A second  boundary condition is that the 
pressure at $r_{ext}$ be
$p\!=\!p_{ext}$.  The two quantities $\dot m_{ext}$ and $p_{ext}$ are, like 
$q$, free parameters of the model. The third boundary condition is provided 
at the free boundary $r\!=\!r_s$ (`sonic radius'), defined as the radius at 
which  $u=-\sqrt{5p/3\rho}$, where the derivatives of the 
unknown functions $\rho, u, p$ are required to be finite. 
This latter condition implies 
\be 
   \rho^2\Lambda(T)\left[1+q\left(2-\frac{{\rm d}\ln\Lambda}{{\rm d}\ln T}
   \right)\right] + \frac{3}{2}\rho u\left(-\frac{{\rm d}\Phi}{{\rm d}r} +
   2\frac{u^2}{r}\right) - \alpha\rho_\star(r)(\epsilon_{inj}+2u^2) = 0 
\label{sonic}\ee
at $r=r_s$. 

The integration is started at a guess for $r_s$, where the mass flux is 
specified in terms of the parameter $\epsilon$, defined as
\be
   \epsilon = \frac{ \dot m (r_s) }{ \alpha \left[M_\star (r_{{ext}}) 
   - M_\star (r_s) \right] + \dot m_{ext}},
\label{epsilon}\ee
which is the gas mass fraction at the sonic radius $r_s$ which has `survived' 
the decoupling process. In the case $q=0$, Eq. (\ref{mass}) is integrated
trivially, with the parameter  $\epsilon$ equal to 1. In the opposite limit of
large $q$,  $\epsilon$ becomes vanishingly small. The values of $r_s$ and 
$\epsilon$ are varied and the integration iterated until the mass flux 
$\dot m$ and the pressure $p$ at $r\!=\!r_{ext}$ are found to match the 
specified values of $\dot m_{ext}$ and $p_{ext}$. Alternatively, when the 
value of $r_s$ is seen to shrink below 0.1pc, so that $\epsilon$ also 
becomes very small, a fully subsonic solution is looked for. The integration 
is then carried inward from $r_{ext}$ starting with a guess for 
$\rho(r_{ext})$ which is improved, for fixed $\dot m_{ext}$ and $p_{ext}$, 
until the central mass flux becomes vanishingly small.  The numerical code 
employed (Toniazzo 1993) uses a double shooting  integration scheme with a 
variable-step fourth-order Runge-Kutta integrator. In the case $q=0$, a 
simple shooting method from $r_s$ outward is found to be viable. When 
$q\neq 0$, such simple shooting method becomes unstable. 

The models illustrated in \S 4 have been chosen with the following procedure. 
For a given galaxy, several models with different parameters have been 
calculated, with the goal of fitting the emission profile and of
matching the value of the integrated luminosity. At a second stage, 
attention was given to the published constraints on the temperature. The 
models are quite sensitive to the values of $q$ and $\dot m_{ext}$, and less to 
$p_{ext}$. 

The code has been tested by checking against the results of BPS, for the 
$q=0$ model of NGC 4472, and the results presented by Sarazin \& Ashe (1989).

\noindent
(v)     {\it Properties of the $q$--models}

In the construction of the ``$q$-models'', we may take advantage of the fact 
that the values of $p_{ext}$ and $\dot m_{ext}$  
have a definite `plausibility' range. It should be stressed that $p_{ext}$ 
and $\dot m_{ext}$
influence only the outer profiles, while for radii
smaller than the optical radius $R_e$ the major role is played by $q$, by the
adopted  gravitational potential $\Phi$ (which in particular determines the 
central temperature, in the transsonic part of the flow), and by the 
adopted supernova rate.  
 
For a given distance to the observed object, the measured X-ray 
surface brightness and temperature at large radii suggest a 
value for the pressure of the gas $p_{ext}$. In particular, {\it Rosat}
data may be able to constrain this quantity for many objects (see, e.g., 
Trinchieri et al. 1994). For models where $q$ plays a significant role, the
amount of gas accreted from the  cluster or from the group can be estimated as
follows. The mass continuity equation (Eq. \ref{mass}) evaluated at  $r_s$
can be
rewritten as:  
\be
 \frac{\dot m_{ext}}{\alpha M_\star} = \frac{q \left[{\it D}(r_{{ext}}) -
{\it D}(r_s)\right]}{(1-\epsilon)~\alpha M_\star} -\delta,
\label{cont}\ee
with $\delta \approx 1$. Since $r_s$ is very small in our models, it may
effectively be replaced by 0, so that the $D$--terms are related to the 
{\it total} radiated power 
\be 
L_t = \int_{0}^{r_{ext}}\rho^2
\Lambda \left[1 + q c(T)\right] 4\pi r^2  {\rm d}r.  
\label{totluminosity}\ee
Then the continuity equation yields the following estimate for the accretion
rate 
\be
\frac{\dot m_{ext}}{\alpha M_\star} \approx \frac{1}{1-
\epsilon}(\frac{0.7 L_t}{L_X}) \left\langle \frac{4 q c(T)}{1 + q c(T)}(\frac{1
{\rm keV}}{k T})\right\rangle\frac{(L_X/10^{41} {\rm erg~s^{-1}})}{(L_B/10^{10}
{\rm L_{\odot}})} - 1,
\label{mdotestimate}\ee
where the angular brackets denote average over the {\it total}
(bolometric) emission. Both the numerator and the denominator in 
Eq.(\ref{mdotestimate}) go to zero when $q\rightarrow 0$, so this equation is 
not useful in this limit. In such limiting case 
$\dot m_{ext}/\alpha M_\star$ has to be estimated directly from the energy
equation that can be derived by integrating Eq.(\ref{enthalpy}). In practice,
all the models selected in the following are in the regime of finite $q$, with 
$q\geq 0.1$ and $\epsilon < 0.1$.

The quantity $L_X$ denotes the power radiated in the relevant X-ray band. 
Following Sarazin \& Ashe (1989),
this can be obtained from the specific emission
\be
b_X(r)=\rho^2\Lambda_X(T)\left[1+
q c(T)
\frac{\Lambda(T)}{\Lambda_X(T)}
\int^T_0\frac{\Lambda_X(T')}{\Lambda(T')}\frac{{\rm d}T'}{T}\right],
\label{emissivity}\ee   
where  $\Lambda_X$ is the emission in the {\it Einstein} $0.5-4.0$ keV energy
band. The emission  spectrum is assumed to be that computed by  Raymond \& Smith
(1977) for a thermal plasma with ``cosmic'' (Allen 1973) abundances. Note the
$q$--dependence of $b_X(r)$, which thus includes the contribution of the
cooling gas that gets decoupled from the hot phase. Therefore, the fraction
of power radiated in the X-ray band $L_X/L_t$ for the model of interest depends
both on the temperature distribution and on the value of $q$. For the selected
models of the following \S 4, this fraction ranges from 0.57 to 0.70.   

If the heating due to supernovae dominates over 
other energy sources, the close similarity of the 
light profile and the X-ray emission profile in the inner parts of X-ray 
bright galaxies displayed by existing {\it Einstein} HRI data is found to occur
naturally in models characterized by $q\gg 1$ (Sarazin 1990). In practice, 
the value of $q\approx 1.3$ turns out to mark the transition to this
asymptotic behavior. As will be shown below, good fits are obtained to the
relevant profiles for $q\approx 0.5$.  


\vspace{2,5truecm}
\noindent
4. THE MODELS\\
\noindent

Three sets of models are presented, corresponding to three different
assumptions on the SNIa rate: $\vartheta_{SN} = 1/16, 1/4$,  and $1.1$. 
Tables 3, 4, and 5 summarize the main global parameters of the models in each
case.

For each model we compute several profiles. Some profiles are specifically
produced for comparison with the observations (see Figs. 2, 3, and 4
corresponding to the three different assumptions on the SNIa rate). The model
X-ray  surface brightness $\Sigma_X(R)$ at projected  distance $R$ is
computed by
integrating the quantity $b_X$ of Eq.(\ref{emissivity}) over the  line-of-sight
line element $d\ell$. For those cases where comparison is made with {\it
Einstein} IPC and {\it Rosat} PSPC data, the surface brightness profiles are
convolved with gaussians of  110'' and 60'' FWHM respectively. The fit process
to the observed surface brightness data is especially aimed at reproducing the
{\it shape} of the observed profiles. The data points shown in the Figures are
treated by allowing for a modest adjustment of the conversion factor from count
rates (given by the original data sources) to the physical units. In the
following, when we will talk about "overbright" models we will mean those for
which the model X-ray luminosity exceeds significantly value for $L_X$ of Table
1.  

The projected emission temperature is computed from
\be
T_X(R)=\frac{\displaystyle
{\Large\int}d\ell ~\rho^2\Lambda_X(T)T\left[1+
q c(T)
\frac{\Lambda(T)}{\Lambda_X(T)}
\int^T_0\frac{\Lambda_X(T')}{\Lambda(T')}\frac{{\rm T'd}T'}{T^2}\right]}
{\Sigma_X(R)},
\label{projtemp}\ee
which is an emission-weighted temperature of the gas. [Note that the
weight entering the integral over the line-of-sight slightly differs from
$b_X(r)$ in the term that describes the contribution of the decoupled gas.] In
the Tables we report for each model the value of an emission averaged (i.e.,
based on $\Sigma_X(R)$, over the whole galaxy) temperature $\langle kT \rangle$.
In general, this latter quantity is different from the temperature that could 
be derived from  a best--fit to the emitted radiation by a Raymond spectrum. 
The result of this fit depends on the sensitivity of the instrument as a 
function of photon energy; in particular, it will be biased towards the peak 
of the detector quantum efficiency. Figure 5 shows  more in detail the 
emission profiles of the three brightest objects, comparing the results of 
models obtained for different assumed SNIa rates.  The profiles of some 
physical quantities (temperature $T(r)$, particle density 
$n(r) = \rho(r)/ \mu m_p$ and mass flux $\dot m (r)$), describing the 
intrinsic properties of the models, are shown in Fig. 6.

\vspace{.6cm}
\noindent
(i) {\it Global properties of the selected models}

If we exclude the case of NGC 1404, which turns out to be a peculiar object
in this modeling context (see comments below), the selected models are 
characterized by a fairly uniform value of $q$ which decreases smoothly when 
the assumed value of $\vartheta_{SN}$ is increased.

The model accretion rates $\dot m_{ext}$, which are smaller for higher 
assumed values of $\vartheta_{SN}$, show a wide variation from galaxy to 
galaxy. The models for NGC 4374 all have a nominal value of 0.1 $\msolar/\yr$. 
For NGC 4472, $\dot m_{ext}$ ranges from 1 $\msolar/\yr$ to 0.3 $\msolar/\yr$. 
The latter value refers to the overbright model with $\vartheta_{SN}=1.1$; in 
this case, a lower accretion rate would not be compatible with the existing 
surface brightness and temperature data at large radii.  The accretion rates 
in NGC 1404 are not large, but $\dot m_{ext}$ is significant when the adopted 
supernova rate is small. In contrast, for any of the assumed choices of 
$\vartheta_{SN}$, most of the gas present in NGC 4636 is found to originate 
from the outside (about 67\% for the lowest value of the supernova rate). The 
energetics of NGC 1399 appears to be dominated by a substantial accretion of 
external gas.
 

The values of the external pressure $p_{ext}$ identified by
our modeling procedure are on the high side, ranging from
$4\times10^3\kelvin~{\rm cm}^{-3}$ to  $1.5\times10^4\kelvin~{\rm cm}^{-3}$. 
The latter value refers to NGC 4374 for a high assumed supernova rate; lower 
values of $p_{ext}$ would also give reasonable models for this galaxy. In 
general, these relatively high values of $p_{ext}$ are produced in order to 
fit a fairly flat surface brightness profile in the outer parts and are 
consistent with the overall increase in the temperature profiles in the outer 
regions suggested by the existing data (see also B\"{o}hringer et al 1994).

We should stress that the high value of $\vartheta_{SN} = 1.1$ suggested by
van den  Bergh \& Tammann (1991) would imply that the energy input is
dominated by supernovae. This explains why the corresponding selected models 
tend to be characterized by larger pressures $p_{ext}$ (especially for the
objects with small $r_{ext}$) and by smaller accretion rates  $\dot m_{ext}$.  
Note that, in the case of high supernova rate, the selected models for
NGC 4374 and NGC 4472 are overbright, by factors of 2.4 and 1.4 respectively. 

The mass of the gas $M_{gas}$ recorded in the Tables gives the integral of
$\rho$ out to $r_{ext}$. The numbers here are found to be quite large compared
to previous estimates (see Forman et al. 1985, Thomas et al. 1986). This is 
particularly true for NGC 1399 and NGC 4636, for which the mass of the hot gas 
is of the same order as that of the stellar component. One obvious worry is
that, under these conditions, the self-gravity of the gas should be
incorporated in the model. Fortunately, the models do have sizable amounts of 
dark matter, so that even in the worst case (model $\vartheta_{SN} =1/16$ for 
NGC 1399) the gas actually makes up no more than  6\% of the {\it total} mass 
inside $r_{ext}$ (this percentage drops to 3\% for NGC 4636, 2\% for NGC 4472 
and NGC 1404 and less than 1\% for NGC 4374).

NGC 1404, as might have been anticipated, turns out to be somewhat peculiar.
Its peaked emission profile and its large $L_X/L_B$ require a small  value for 
the parameter $q$ (0.1), independently of the assumed supernova rate (however, 
the $q=0$ models are ruled out). The cooling flow models for this object also 
stand out because of their  relatively flat $\dot m$ profiles (see further 
discussion in item (iii) below). It appears that the characteristics of this 
galaxy mostly result from the interaction with its environment, dominated by 
NGC 1399. From the point of view of the  observational input, one has to face 
a non--trivial subtraction problem, since the galaxy is embedded in the bright 
halo of NGC 1399. The models adopted are clearly oversimplified with respect 
to the actual physical situation. Still we find no particular  difficulty in
fitting the emission  profile of NGC 1404. In view of its shallow potential 
well and of the constraints posed on $p_{ext}$ by the apparent proximity to 
NGC 1399, a relatively low average X-ray emission temperature ($\approx 0.6$ 
keV) is predicted for the hot gas. 

>From the properties of the environment identified by our models, it turns out 
that the intergalactic medium in Virgo and Fornax should be characterized by 
relatively high densities, so that ram--pressure stripping may be expected 
to be efficient. This is indeed suggested by White and Sarazin (1991), who 
find a correlation between the dispersion in $L_X$ for a given $L_B$ and the 
local number density of galaxies. On the other hand, the effectiveness of 
ram--pressure stripping largely depends on the velocity of the galaxy relative 
to the medium (Gaetz et al. 1987). 
In particular, for typical values of $\rho(r_{ext})$ in our models 
($10^{-3}$ amu), the condition for efficient stripping (see Eq.(10) in White 
and Sarazin 1991) does not seem to be satisfied for velocities below $500$ 
km/s. Indeed, the X-ray isophotes of NGC 1399 and NGC 4636 are fairly round. 

After submission of this paper, some interesting results from the {\it ASCA} 
mission have come to our attention (Loewenstein et al. 1994; Mushotzky et al. 
1994), which may provide an additional way to discriminate our models, based on 
their average temperatures $\langle kT \rangle$ (see column 8 of Tables 3--5). 
For NGC 4636 the {\it ASCA} temperature measurements are consistent with 
those of Trinchieri et al. (1994) (based on {\it Rosat} data), which are 
accounted for in our models. For NGC 4374 and NGC 1404 the values of 
0.74 and 0.75 keV seem to point in the direction of the choice  
$\theta_{SN} = 1/4$. For
these two galaxies, a quick investigation has shown that models can be
produced, characterized by average temperatures consistent with the {\it ASCA}
values; in particular, in the case of NGC 1404 a model with $q = 0.2$,  $\dot
m_{ext} = 0.9 \msolar /yr$, $p_{ext} = 2.2 \times 10^4 \kelvin~{\rm cm}^{-3}$,
and $\theta_{SN} = 1/4$ gives $\langle kT \rangle = 0.79$ keV with a reasonable
emission profile.

\vspace{.6cm}
\noindent
(ii) {\it Surface brightness and temperature profiles}

For four objects in our sample,  both the emission and the temperature 
profiles compare well with the available {\it Einstein} IPC data. The case of 
NGC 4636 stands out because the photometric fit is unsatisfactory for both IPC 
and {\it Rosat} PSPC data. While the latter data are closer to the model 
profiles, they also show a clear lack of circular symmetry; still the 
temperature profile is sufficiently well reproduced. The impact of the assumed 
supernova rate $\vartheta_{SN}$ is best noted in the innermost parts of the 
profiles, as shown in Fig. 5 (logarithmic scale) in comparison with HRI data. 
In the low $\vartheta_{SN}$ case, we find an excess of emission with respect 
to the innermost {\it Einstein} HRI data points for the galaxy NGC 4472; when 
a higher supernova rate is considered, the  fit to the high resolution HRI 
data in the central regions definitely improves.  No  important changes in the 
emission and temperature profiles on the large scale  are noted. For the case 
of high supernova rate ($\vartheta_{SN} = 1.1$), the model temperature profile 
of NGC 4636 is not fully satisfactory  when confronted with {\it Rosat} PSPC 
data; this might be improved by allowing for a larger  $\dot m_{ext}$, but in 
such a case the model would be overbright.

Figure 5 also shows the behavior of the best--fit model for the case $q = 0$
considered by BPS. This gives convincing evidence that the $q$--models are
indeed able to remove almost completely the objection raised against simple
cooling flow models in relation to the predicted emission profiles.

\vspace{.6cm}
\noindent
(iii) {\it Intrinsic profiles}

In Fig. 6 we summarize the properties of the intrinsic profiles that
characterize the selected models.   

For the three galaxies (NGC 1399, NGC 1404, and NGC 4636) with the highest
values of  $\dot m_{ext}/\alpha M_\star$, the accretion rate
($\dot m$) profiles are monotonically increasing; for NGC 1404 and NGC 4636,
characterized by the lowest model values of $q$, such profiles are relatively
flat. For NGC 4374 and NGC 4472, the accretion rate is non--monotonic in the
vicinity of $r_{ext}$.

>From Fig. 6 one can see that our models are computed down to very small 
radii ($ r < 10$ pc). The models are probably
unphysical at such small galactocentric distances, but the profiles are
shown in order to display the internal properties of the models and of the
boundary conditions that have been adopted.

\vspace{.6cm}
\noindent
(iv) {\it Consistency of the steady-state cooling flow description}

In Fig. 7 we present the characteristic timescales for the models selected 
under the assumption of $\vartheta_{SN} = 1/4$. In each frame the three
curves represent, as a function of galactocentric radius, the cooling timescale
$\tau_{cool}$ (see Eq.(\ref{thintime})), the flow timescale (i.e., the time 
necessary for a fluid element to reach the center by moving at the fluid 
velocity $u(r)$), and the sound crossing time (i.e., the time necessary for a 
signal moving at the speed $u_s = -\sqrt{5p/3\rho}$ to reach the center). 
Models characterized by smaller values of $q$ (as is the case of NGC 1404) 
tend to have smaller values of the flow timescale at $r_{ext}$. 

>From Fig. 7 one can see that the flow timescale actually
exceeds the Hubble  time in the outer parts of the flow (dashed lines). On
the other hand, while the cooling time is even longer (solid lines),
pressure  equilibrium is guaranteed by the short ($\sim 10^8\yr$, dotted lines)
sound  crossing time. We must conclude that the steady-state scenario is,
strictly speaking, unjustified at very large radii; the calculated profiles are
shown there also to describe the properties of the models at the outer
boundary.

\vspace{.6cm}
\noindent
(v) {\it Some general trends from the selected sample of galaxies}

>From the above discussion we find that, for a very low value of the supernova 
rate, the profiles are best fitted by models characterized by high values of 
$q$; still the central parts of the surface brightness profiles of the models 
are too steep. In contrast, based on $\theta_{SN}=1.1$, the model temperature 
in the outer regions of NGC 4636 seems to be higher than that observed, while
the  model X--ray luminosities for NGC 4472 and NGC 4374 are in excess of the 
observed values by factors 1.4 and 2.4 respectively. The intermediate case 
($\theta_{SN}=1/4$) is probably the one to be preferred, if one assumes a 
strictly constant rate from galaxy to galaxy, although the innermost data 
point for the HRI profile of NGC 4472 is not well accounted for. 


In our limited sample of galaxies, two potentially interesting correlations 
have been noted and are shown in Fig. 8. The right frame gives the correlation
between $\dot m_{ext}/\alpha M_\star$ and $(L_X/10^{41} {\rm
erg~s^{-1}})/(L_B/10^{10} {\rm L_{\odot}})$:

\be
\frac{\dot m_{ext}}{\alpha M_\star} \approx a \frac{(L_X/10^{41} {\rm
erg~s^{-1}})}{(L_B/10^{10} {\rm L_{\odot}})} - b.
\label{correlation}\ee

A linear regression gives  $a = 1.51, 1.53, 1.20$ and $b = -0.11, 0.11,
0.40$ (for $\vartheta_{SN} = 1/16, 1/4, 1.1$ respectively). This may be 
compared to the estimate given in Eq. (\ref{mdotestimate}).  
Thus a natural interpretation for the different $L_X/L_B$ values that are
observed in our sample of elliptical galaxies suggests
that, because of dilution, the iron abundance in the hot gas of X-ray bright
ellipticals should anticorrelate with $L_X/L_B$. The opposite
trend would result if the higher $L_X/L_B$ ratios were ascribed to higher SNIa
rates, but this would be hard to justify.  

In this scenario of a ``diluted cooling flow'' one would expect a metallicity 
gradient to be established preferentially for objects having high $L_X/L_B$. 
The decoupling of the gas from the cooling flow would enhance the effect. 
Clearly, a quantitative assessment of the actual dilution factor would require 
a full modeling of the history of the galaxy--cluster interaction, which is 
well beyond the limits of the steady--state analysis adopted in this paper. 
Such a detailed modeling would provide a determination of the size of the 
above mentioned metallicity gradient. At this stage we feel encouraged by the 
qualitative trends reported by Loewenstein et al. (1994) and Mushotzky et al. 
(1994). 

Another possibly interesting trend that is noted is a fairly tight  
correlation between $\dot m_{ext}$ and the ratio $r_h/r_L$, where $r_h$
denotes the half-mass radius of the {\it total} (dark + luminous) mass
distribution of the galaxy. Here the offset of NGC 1404 might be due to the
peculiarities of this object pointed out above. While the X-ray
emission and temperature profiles appear to be of modest significance in
diagnosing the presence of dark halo in galaxies (see BPS), the total X-ray
luminosity might be a significant indicator (see also Bertin et al. 1994). We
recall that no other significant correlations of $L_X/L_B$ with intrinsic 
galaxy properties have been noted (see White \& Sarazin  1991).

\vspace{2,5truecm}
\vspace{3cm}

\noindent
5. CONCLUSIONS
\noindent

We have shown that {\it within galaxy models constrained by radially
extended stellar dynamical data}, a simple spherically-symmetric, steady-state,
cooling-flow  description with gas loss, generally compares well with the
existing data on  X-ray emission profiles and temperatures for a set of bright
elliptical galaxies.  

Three free parameters are needed to specify the model: (i)
the external mass flux $\dot m_{ext}$, which represents accretion of
intergalactic gas and increases the X-ray luminosity of the galaxy, (ii) the
external pressure $p_{ext}$, which depends on the X-ray temperature of the
cluster or group  embedding it and helps confine the gas in the galactic
potential well, and (iii) the  dimensionless coefficient $q$, which 
parameterizes the decoupling of thermally  unstable clouds of cooling gas 
from the homogeneous phase  (Sarazin \& Ashe  1989) and regulates the mass 
deposition rate along the flow. 

Especially the use of the $q$ parameter leads to significantly better fits
to the X-ray emission profiles with respect to the $q=0$ models, while the
derived temperature profiles are consistent with, but only loosely constrained
by, the existing data. Still it is evident that in some cases the assumption of
spherical symmetry is an oversimplification; in addition, we must be aware
that the steady-state condition is not fulfilled in the outer regions.  

We have found  that a value for the supernova rate in elliptical galaxies 
consistent with the estimate of Cappellaro et al. (1993) is favored by our 
models. Confining pressures of $p_{ext}\sim4\div15\times10^3\kelvin$ cm$^{-3}$ 
and  significant accretion rates of external material, up to $4\msolar/\yr$, 
are suggested by our models. Finally, we have argued that the value of 
$L_X/L_B$ should anticorrelate with the iron abundance in the gas inside 
ellipticals. 

\vspace {1cm}

\noindent

We would like to thank L. Ciotti, R. P. Saglia, C. L. Sarazin, and G. 
Trinchieri for valuable suggestions and for their collaboration. This work has
been partially  supported by ASI (under contract ASI 94 RS 94 -- 202/3 FAE) of
Italy.

\newpage
%
%
\begin{center}

{\large Table 1. Optical and X-ray properties of the selected objects.}

\begin{tabular}{cccccccc}
\\
\hline
 NGC    & Type$^{(7)}$  & $D^{(8)}$     & $m_B^{(7)}$   & $R_X$ & log $L_X$
 & $\langle T_X \rangle$ & log$(L_X/L_B)^{(9)}$
\\
        & & (Mpc) & (mag) & (")  & (erg/s) 
& (keV) & 
\\ \hline 
 1399   &   E1P   &  28   & 9.85  & 1200$^{(1)}$ & 42.33$^{(1)}$
 & $\ge 1.1^{(2)}$ & 31.46  \\
        &         &       &       & 120$^{(3)}$  & 41.63$^{(3)}$
 &  1.02$^{(3)}$  & 
\\
 1404   &   E1    &  28   & 10.87 & 300$^{(1)}$  & 41.49$^{(1)}$   
 & $\ge 0.8^{(2)}$ & 31.10  \\
        &         &       &       & 220$^{(4)}$  & 41.36$^{(4)}$
 &--    &
\\
 4374   &   E1    &  27   & 9.98  & 300$^{(1)}$  & 41.16$^{(1)}$
 & $\ge 0.8^{(2)}$ & 30.20  \\
        &         &       &       & 150$^{(5)}$  & 40.84$^{(5)}$
 & 0.9--1.3$^{(5)}$ &
\\
 4472   &   E2    &  27   & 9.09  & 810$^{(1)}$      & 42.06$^{(1)}$  
 & 1.2$^{(2)}$ & 30.73  \\
        &         &       &       & $\sim$200$^{(3)}$& 41.77$^{(3)}$ 
 & 0.92$^{(3)}$ & 
\\
 4636   &   E0+   &  27   & 10.44 & 600$^{(1)}$  & 41.99$^{(1)}$
 & 0.9$^{(2)}$ & 30.99 \\
        &         &       &       & 1080$^{(6)}$ & 42.27$^{(6)}$
 & 0.87$^{(6)}$ & \\
\hline
\end{tabular}
\end{center}

{\footnotesize 
\begin{itemize}
\item[$^{(1)}$] Fabbiano, Kim \& Trinchieri 1992 (0.2--4.0 keV band)

\item[$^{(2)}$] Kim, Fabbiano \& Trinchieri 1992 

\item[$^{(3)}$] Serlemitsos et al. 1993 (0.5--4.5 keV band)

\item[$^{(4)}$] Thomas et al. 1986 

\item[$^{(5)}$] Forman, Jones \& Tucker 1985 (0.5--4.5 keV band). For the 
temperature a 90\% confidence interval is reported.

\item[$^{(6)}$] Trinchieri et al. 1994 (0.1--2.4 keV band). The 
temperature recorded here was computed as an emission weighted average. 

\item[$^{(7)}$] RC3 (de Vaucouleurs 1991)

\item[$^{(8)}$] Faber et al. 1989 

\item[$^{(9)}$] $L_X$ from column 6; $L_B$  in solar units from SBS 
(see Table 2)
\end{itemize}}


\vspace{1,5truecm}
\begin{center}
{\large Table 2. Stellar--dynamical models (from SBS).}

\begin{tabular}{cccccccc}
\\
\hline
 NGC    &      $L_B$      & $r_L$ & $r_L$ &     $M_{\star}/L_B$    
 &    $M/L_B$        & $r_D/r_L$ & $\sigma_\star(0)$
\\
        & $(10^{10}{\rm L}_\odot)$&  (")  &  (kpc)  & 
$({\rm M}_\odot/{\rm L}_\odot)$ & $({\rm M}_\odot/{\rm L}_\odot)$ & & (km/s) 
\\ \hline 
%
%
 1399 &  7.5  & 136 &  18.4  &  8.0  &  58.0  &  6.2  & 250 
\\
 1404 &  2.5  &  90 &  12.6  &  4.6  &  12.0  &  1.0  & 200
\\
 4374 &  9.2  & 179 &  23.5  &  5.3  &  14.2  &  1.0  & 308
\\
 4472 & 21.2  & 173 &  22.0  &  5.0  &  13.4  &  2.8  & 316    
\\
 4636 &  9.9  & 215 &  28.1  &  8.0  &  38.6  &  5.9  & 227
\\ \hline
\end{tabular}
\end{center}

\newpage

%
%

{\large Table 3: Properties of the selected models ($\vartheta_{SN} = 1/16$)}

\begin{center}
\begin{tabular}{ccccccccc}
\hline
 NGC  &  $q$  & $\dot m_{ext}$ & $p_{ext}$ & $r_{ext}$ & $M_{gas}$ 
& log($L_X$) & $\langle kT \rangle$ & $\dot m_{ext}/\alpha M_\star$
\\
      &     &  (M$_\odot$/yr)   & (10$^3~^0$K/cm$^3$) & (kpc) 
& (10$^{10}$M$_\odot$) & (erg/s) & (keV) & 
\\ \hline
 1399 & 0.8 &  4.4  &   8.2 &  200  &  19.71  &  42.31 & 1.31 & 3.91
\\
 1404 & 0.1 &  0.9  &  10.0 &   50  &   0.69  &  41.41 & 0.52 & 2.40
\\
 4374 & 2.0 &  0.1  &   4.9 &   60  &   0.24  &  41.14 & 0.75 & 0.07
\\
 4472 & 0.8 &  1.0  &  10.1 &  120  &   4.42  &  42.02 & 1.09 & 0.31
\\
 4636 & 0.5 &  3.0  &   4.0 &  160  &   7.95  &  42.00 & 0.84 & 2.02
\\
\hline
\end{tabular}
\end{center}

\vspace{1cm}

{\large Table 4: Properties of the selected models ($\vartheta_{SN} = 1/4$)}

\begin{center}
\begin{tabular}{ccccccccc}
\hline
 NGC  &  $q$  & $\dot m_{ext}$ & $p_{ext}$ & $r_{ext}$ & $M_{gas}$ 
& log($L_X$) & $\langle kT \rangle$ & $\dot m_{ext}/\alpha M_\star$
\\
      &     &  (M$_\odot$/yr)   & (10$^3~^0$K/cm$^3$) & (kpc) & 
(10$^{10}$M$_\odot$) & (erg/s) & (keV) & 
\\ \hline
 1399 & 0.6 &  4.1  &   7.6  &  200  &  19.55  &  42.29 & 1.24 & 3.65
\\
 1404 & 0.1 &  0.6  &   7.6  &   50  &   0.66  &  41.35 & 0.60 & 1.60 
\\
 4374 & 0.8 &  0.1  &   6.3  &   60  &   0.51  &  41.35 & 0.85 & 0.07
\\
 4472 & 0.8 &  0.6  &  12.6  &   120 &   4.55  &  42.01 & 1.22 & 0.19
\\
 4636 & 0.4 &  3.0  &    3.8  &  160  &   8.20  & 42.02 & 0.80 & 2.02
\\
\hline
\end{tabular}
\end{center}

\vspace{1cm}

{\large Table 5: Properties of the selected models ($\vartheta_{SN} = 1.1$)}

\begin{center}
\begin{tabular}{ccccccccc}
\hline
 NGC  &  $q$  & $\dot m_{ext}$ & $p_{ext}$ & $r_{ext}$ & $M_{gas}$ 
& log($L_X$) & $\langle kT \rangle$ & $\dot m_{ext}/\alpha M_\star$
\\
      &     &  (M$_\odot$/yr)   & (10$^3~^0$K/cm$^3$) & (kpc) & 
(10$^{10}$M$_\odot$) & (erg/s) & (keV) & 
\\ \hline
 1399 & 0.4 &  3.3  &   9.9  &  200  &  22.65  &  42.33 & 1.40 & 2.94
\\
 1404 & 0.1 &  0.3  &   8.9  &   50  &   0.65  &  41.33 & 0.67 & 0.80
\\
 4374 & 0.3 &  0.1  &  15.1  &   60  &   1.32  &  41.60 & 1.07 & 0.07
\\
 4472 & 0.3 &  0.3  &  12.6  &  120  &   6.00  &  42.14 & 1.17 & 0.09
\\
 4636 & 0.2 &  1.8  &   4.1  &  160  &   9.14  &  42.06 & 0.80 & 1.21
\\
\hline
\end{tabular}
\end{center} 



\newpage
\noindent
{\bf Figure Legends}\\

\noindent
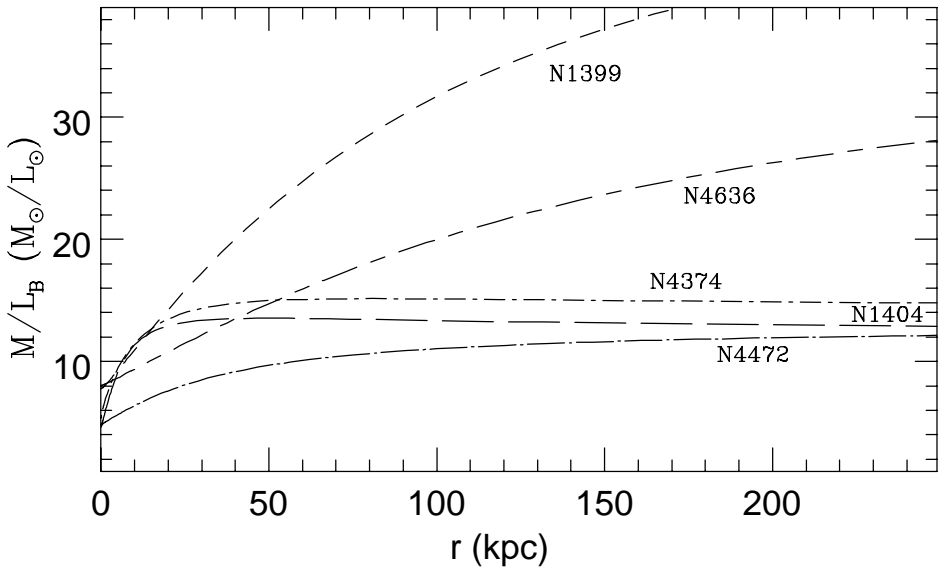
{\bf Figure 1}. Cumulative $M/L_B$ ratios for the best fit stellar dynamical
models (from SBS) of the five elliptical galaxies of the sample. The
models are made of a luminous component of total mass $M_{\star}$ and of
a dark halo (with finite mass) with different spatial distribution, within
a common self-consistent gravitational field. Thus the gradient in the
cumulative $M=M(r)$ reflects the presence of a significant dark halo (see Table
2).


\noindent
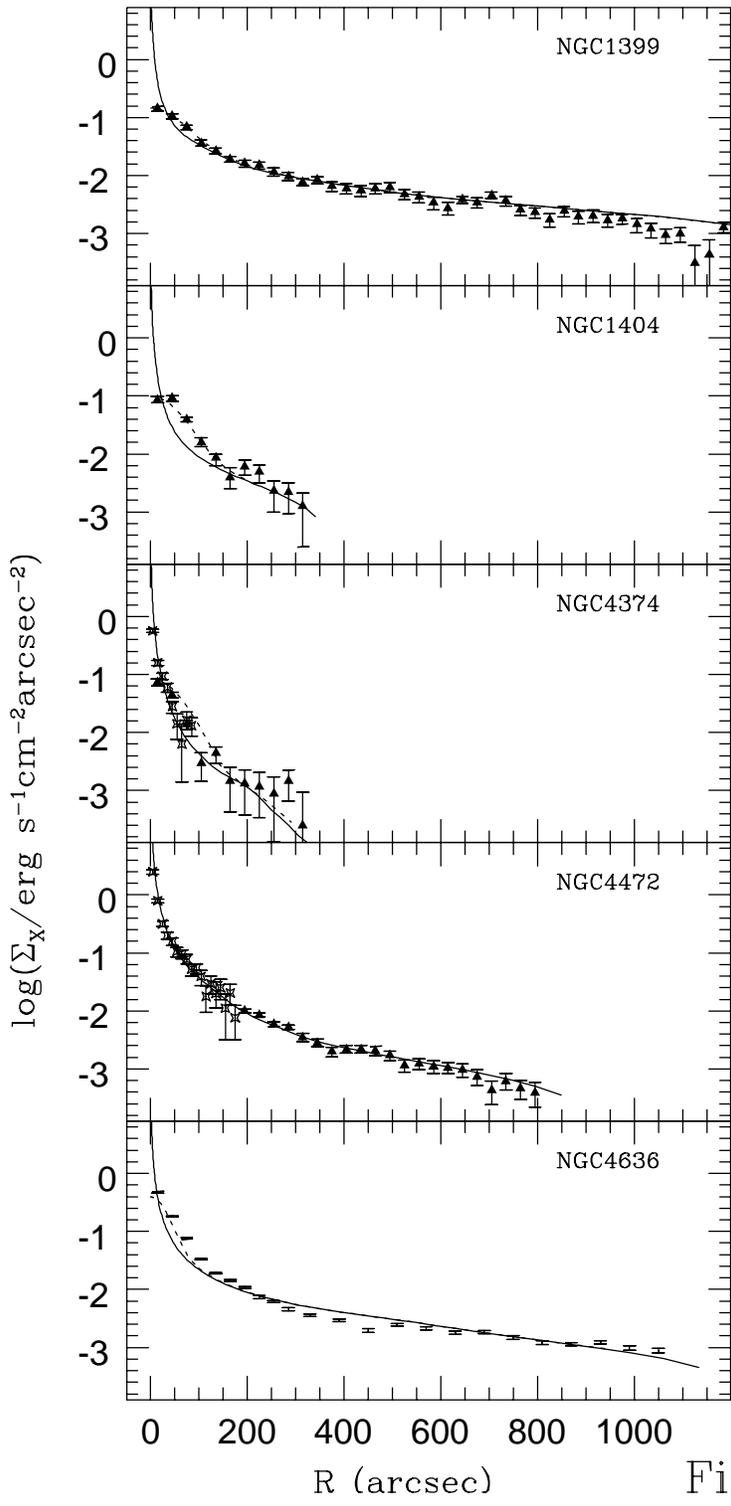
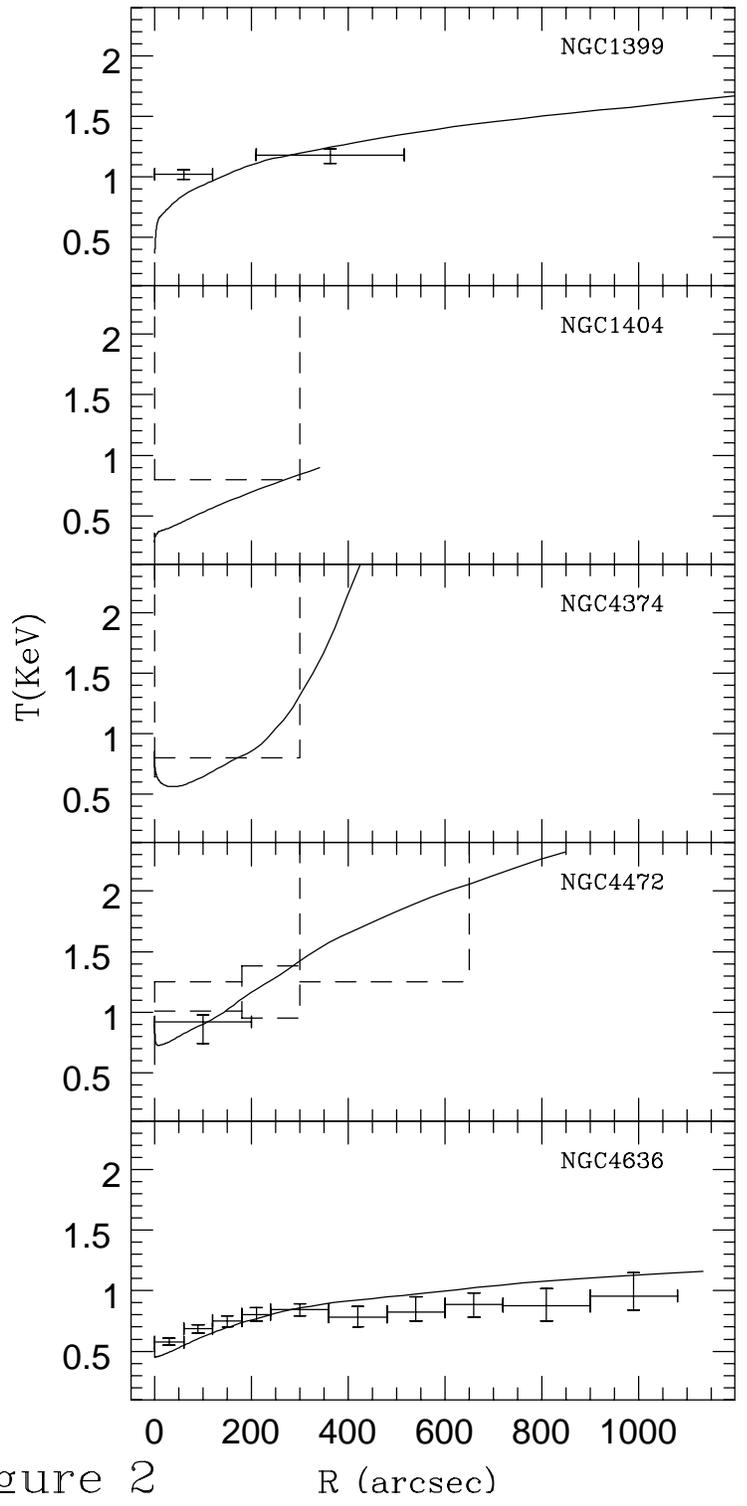
{\bf Figure 2}. X-ray emission and temperature profiles for the
selected models with $\vartheta_{SN} = 1/16$. 
{\it Left frames}: The solid lines are the model X-ray surface brightness
profiles; the dotted lines (shown for all the objects with the exception of NGC
4472) are the result of a convolution with a gaussian PSF, suitable for a
comparison with the IPC or PSPC data (see text).  Triangles indicate {\it
Einstein} IPC data (Fabbiano et al. 1992), crosses are used for the
{\it Einstein} HRI data (for NGC 4472; Fabbiano et al. 1992) and bare errorbars
are {\it Rosat}  PSPC data (for NGC 4636; Trinchieri et al. 1994).  {\it Right
frames}: X-ray temperature profiles. The solid lines represent the projected
emission temperature (see Eq. \ref{projtemp}); the regions outlined
by the dashed lines are the 90\% confidence intervals derived from  {\it
Einstein} IPC data (Kim et al. 1992);  errorbars are the temperature estimates
derived by the {\it BBXRT}  (for NGC 1399 and NGC 4472; Serlemitsos et al. 1993)
and the {\it Rosat} PSPC (for NGC 4636; Trinchieri et al. 1994)  instruments. 


\noindent
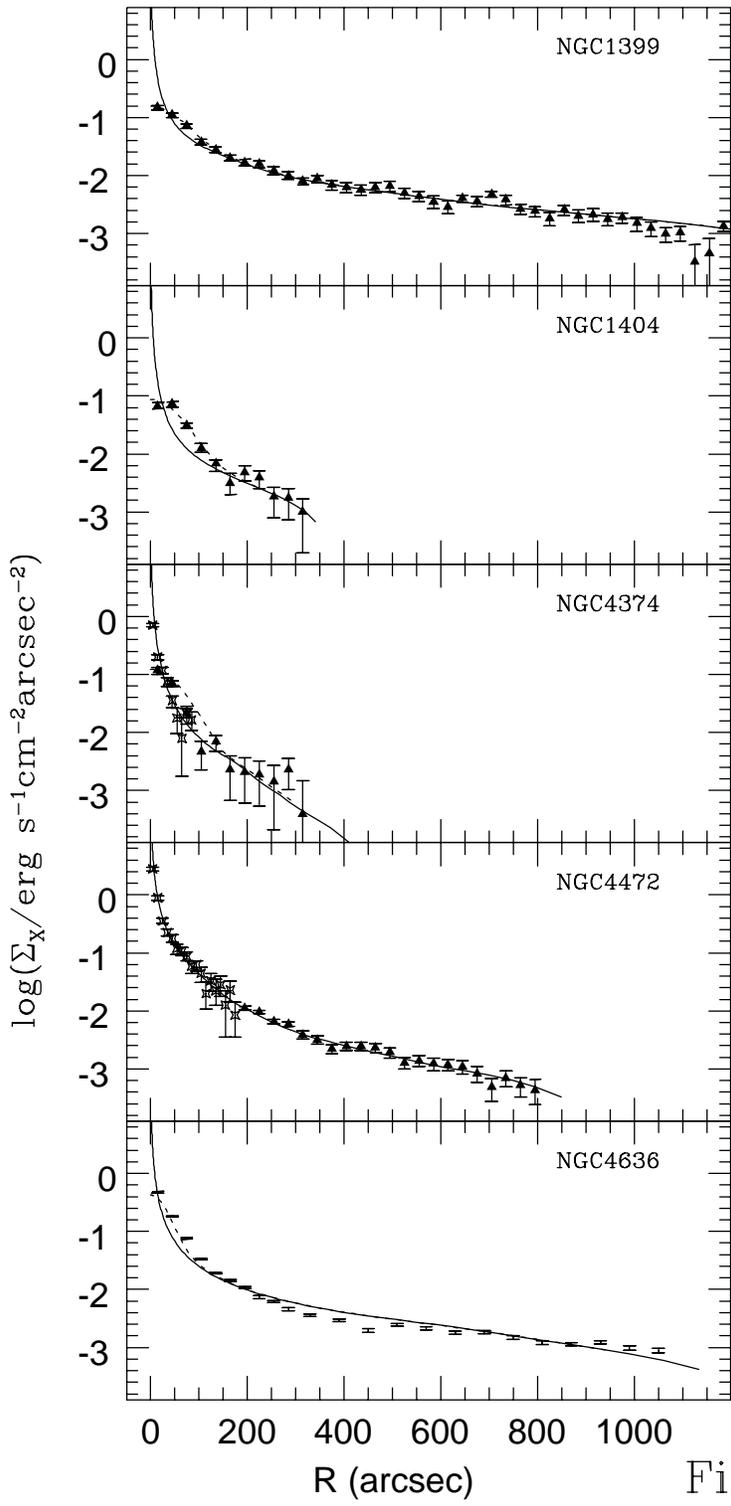
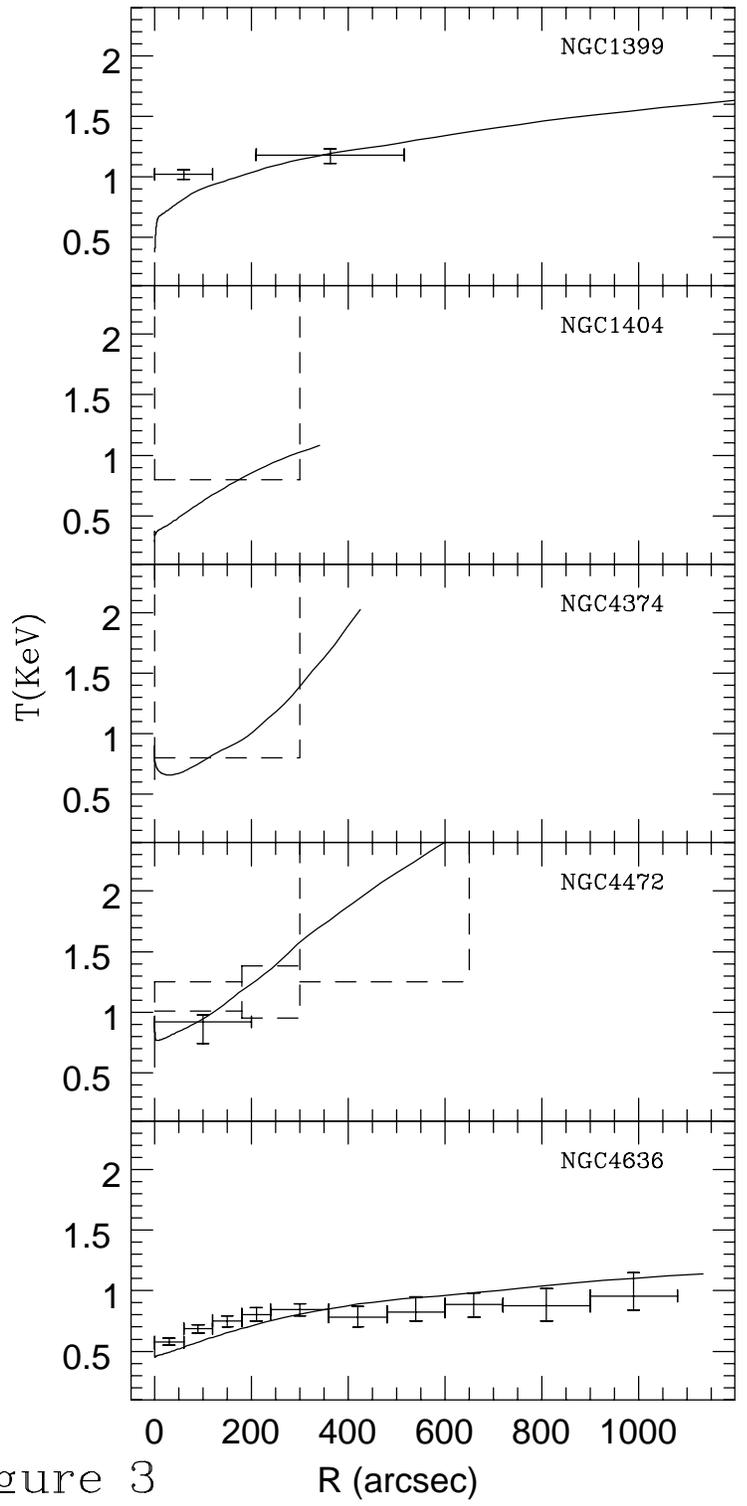
{\bf Figure 3}. Same as Fig. 2, but for $\vartheta_{SN} = 1/4$ models. 


\noindent
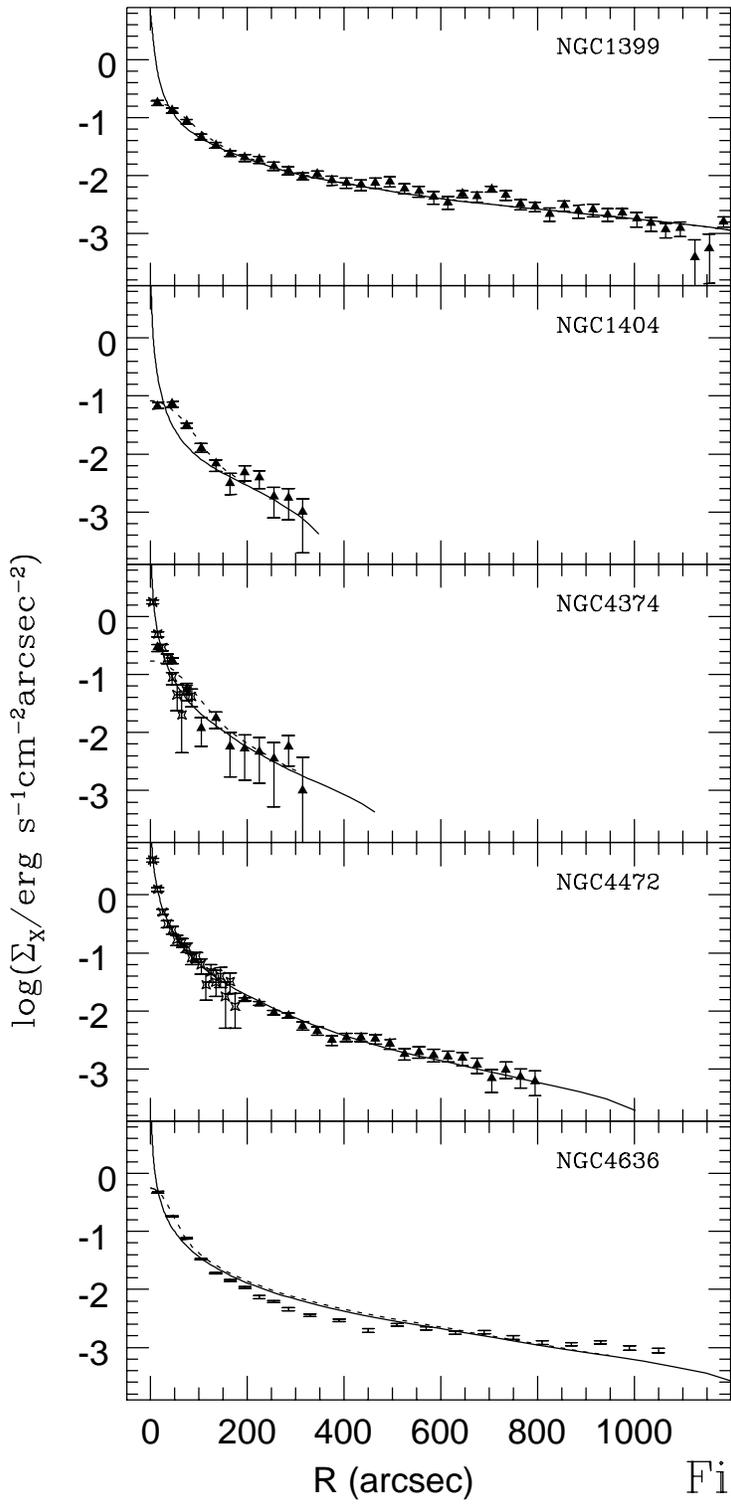
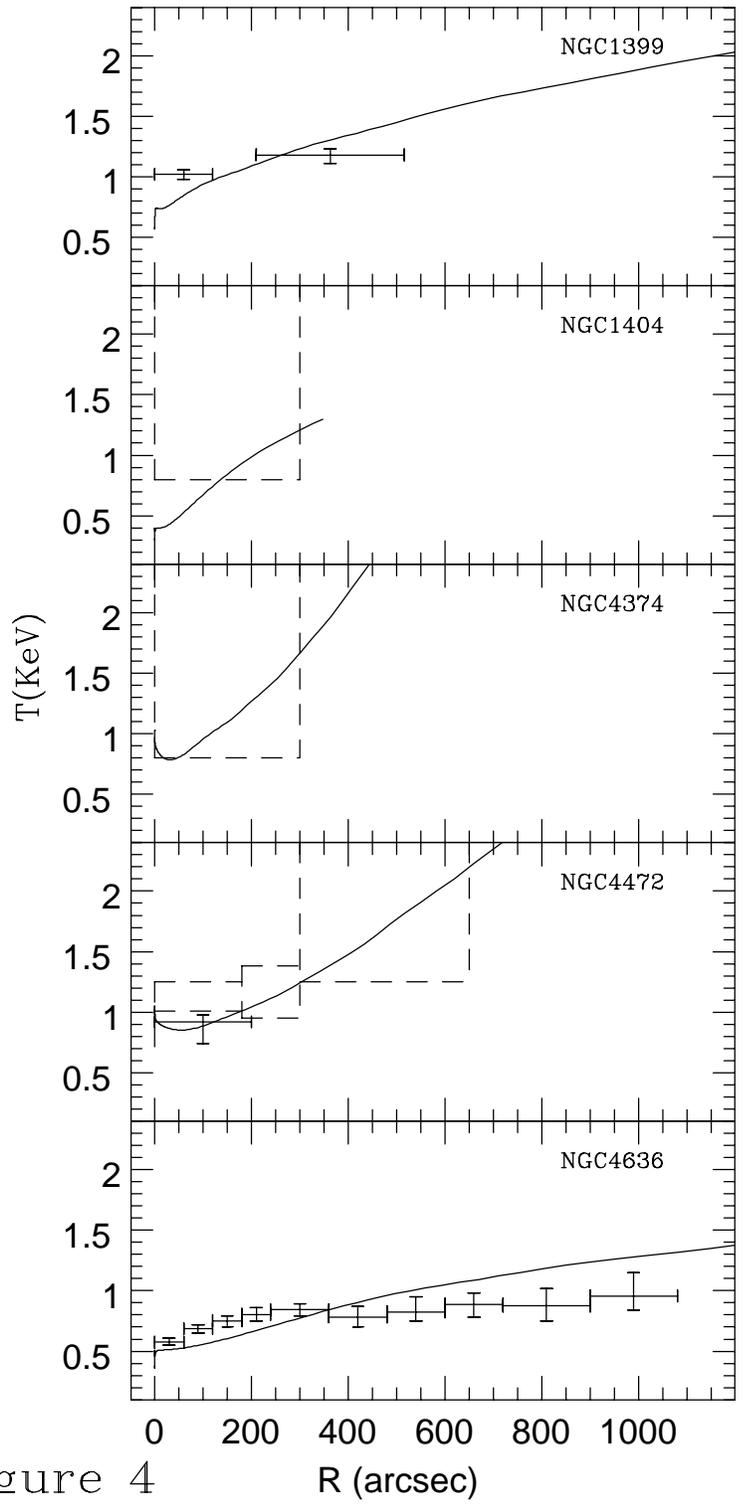
{\bf Figure 4}. Same as Fig. 2, but for $\vartheta_{SN} = 1.1$ models. 


\noindent
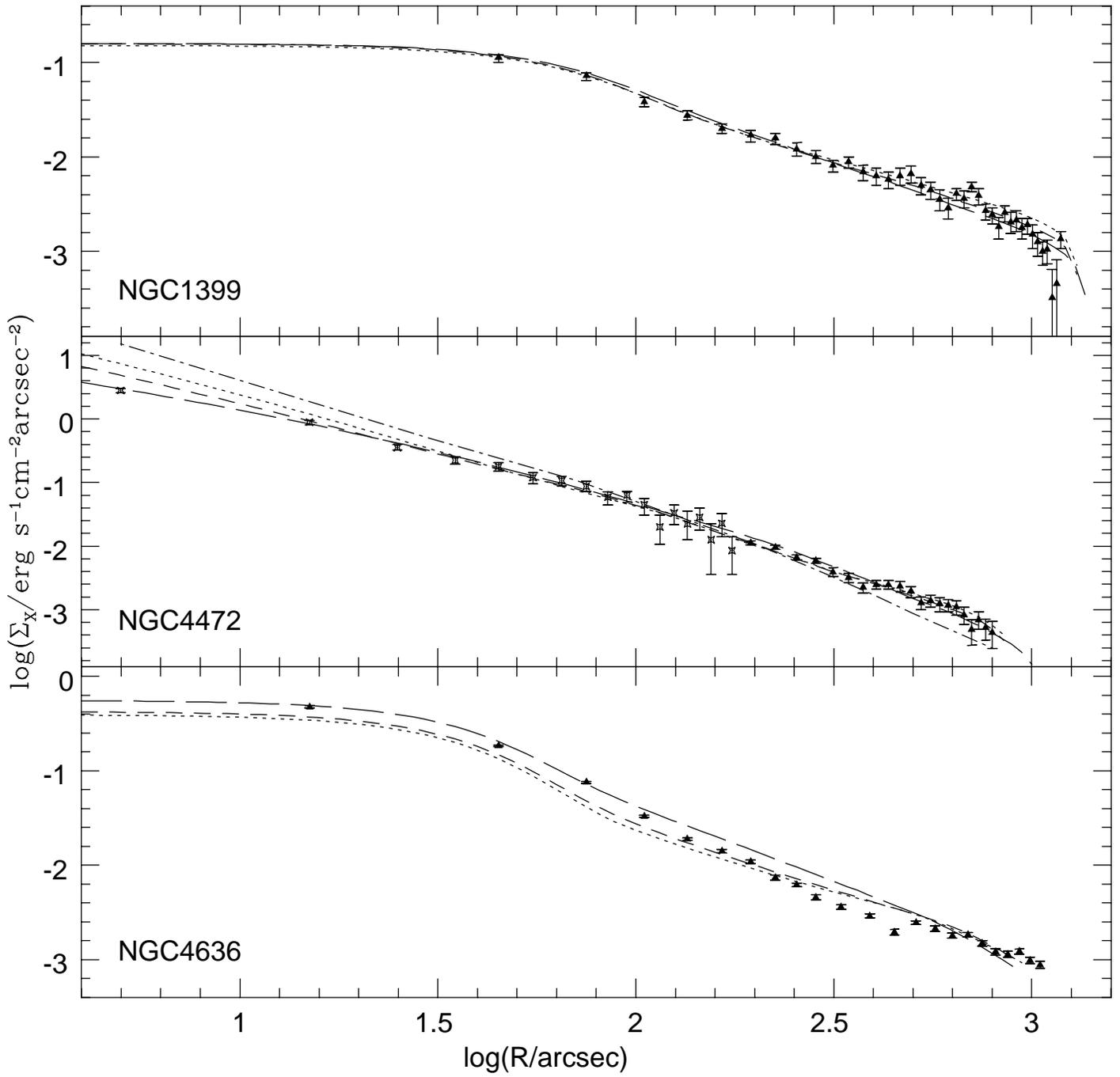
{\bf Figure 5}. X-ray emission profiles for the brightest 
objects showing in detail the behavior in the innermost regions. The
dotted lines refer to models with $\vartheta_{SN} = 1/16$, the
short-dashed lines to  $\vartheta_{SN} = 1/4$, and the long-dashed lines to
$\vartheta_{SN} = 1.1$. Data points are {\it Einstein} IPC and HRI and
{\it Rosat} PSPC data (as in Fig. 2). For NGC 4472 the dashed-dotted line
showing the largest excess of emission in the central region corresponds
to the FF model of BPS (with $q=0$).


\noindent
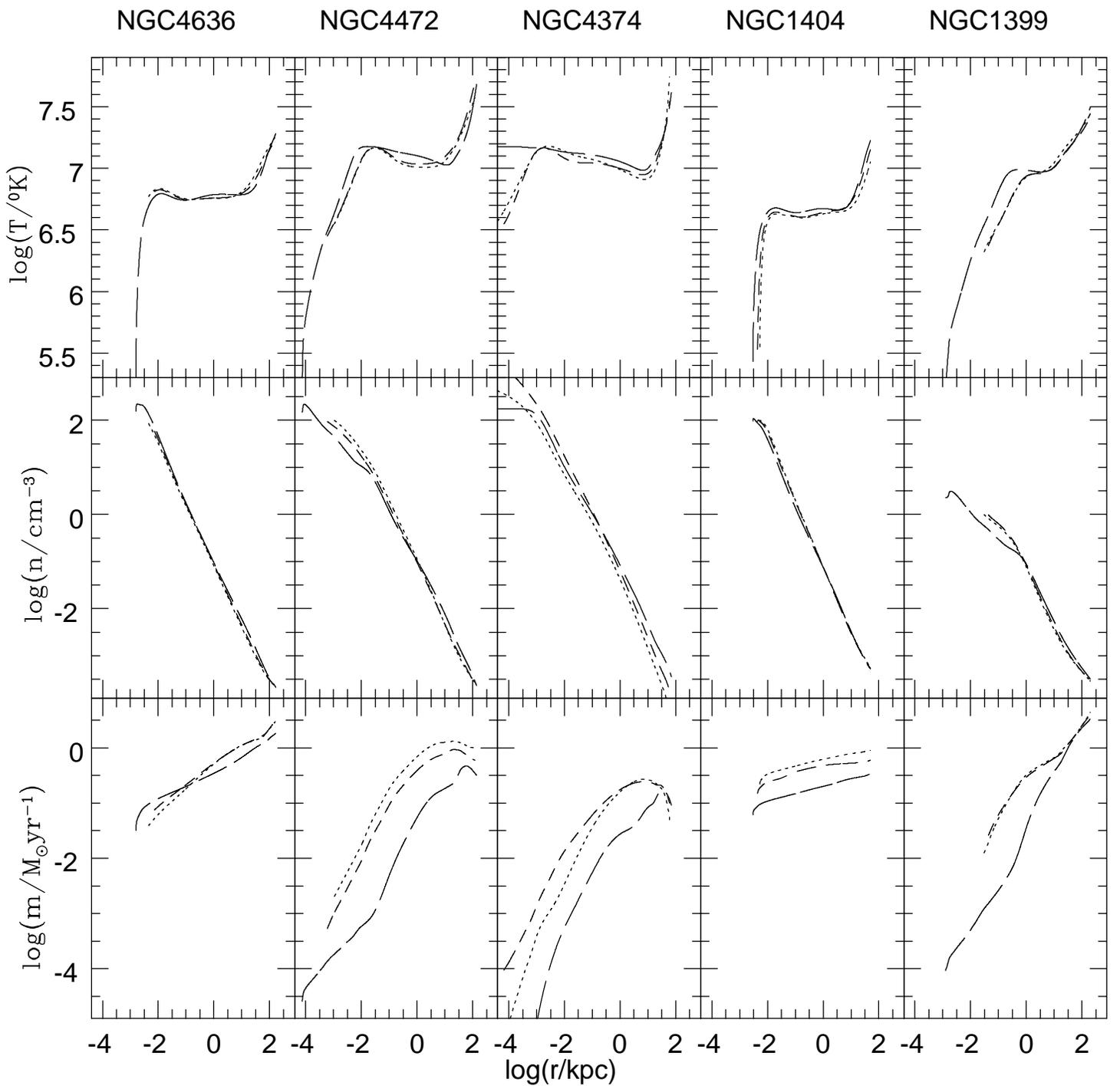
{\bf Figure 6}. Intrinsic properties (temperature, particle density, and mass
flux) of  the selected models.  As in Fig. 5, the dotted lines refer to models
with $\vartheta_{SN} = 1/16$, the short-dashed lines to $\vartheta_{SN} = 1/4$,
and the long-dashed lines to $\vartheta_{SN} = 1.1$.  


\noindent
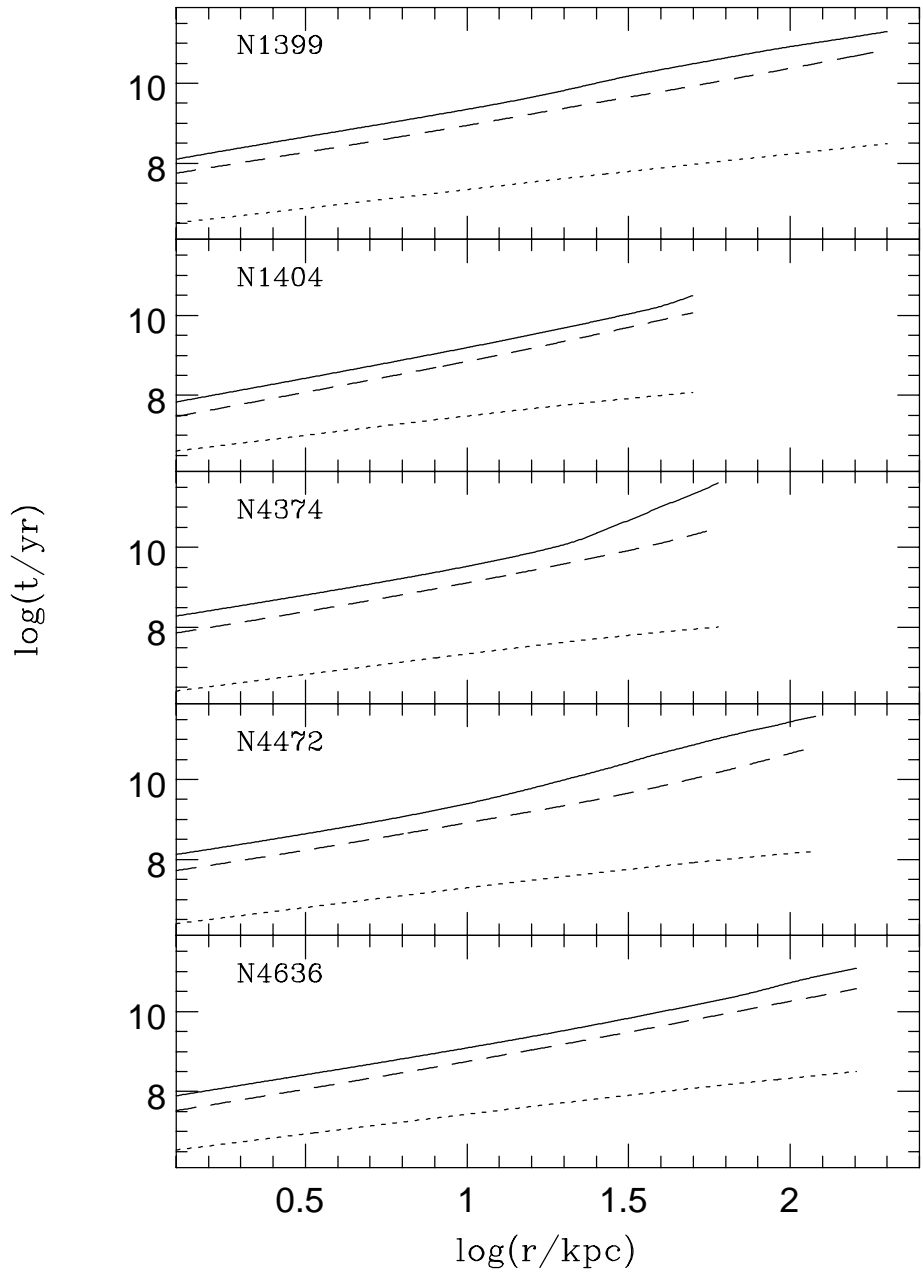
{\bf Figure 7}. Characteristic time scales in the cooling flows for the
selected models with $\vartheta_{SN} = 1/4$. The solid  lines represent the 
cooling times, the dashed lines the flow times, and the dotted lines the 
sound crossing times (see text). 


\noindent
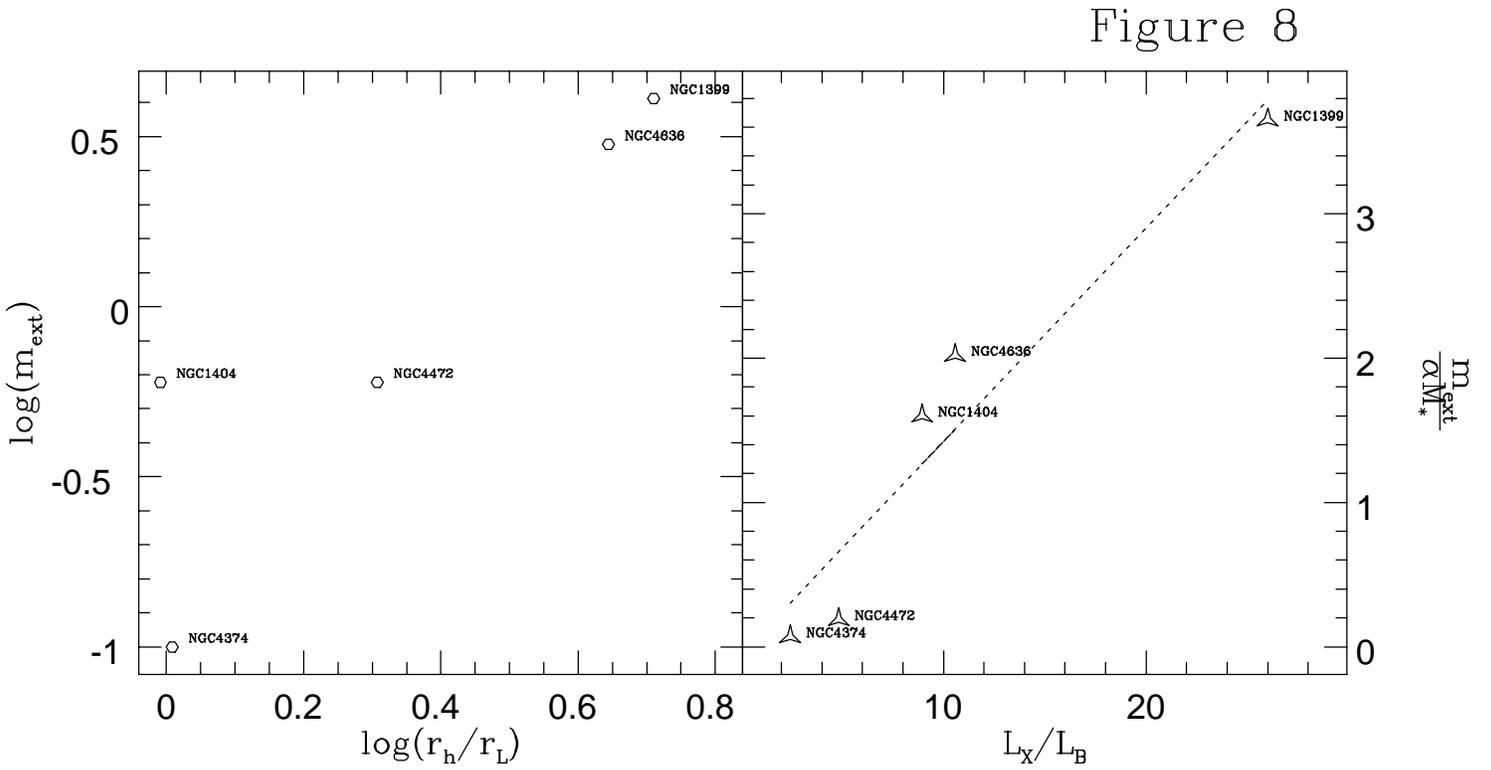
{\bf Figure 8}. Left panel: the mass accretion rate $\dot m_{ext}$ is found
to
be higher for diffuse halos ($r_h$ is the half-mass radius of the {\it total}
(dark + luminous) mass, while $r_L$ is the half-mass radius of the luminous
component). Right panel: scaling of the dilution parameter 
$\dot m_{ext}/\alpha M_\star$ with the relative X-ray luminosity $(L_X/10^{41}
{\rm erg~s^{-1}})/(L_B/10^{10} {\rm L_{\odot}})$. Here the values of the
dilution parameter and of the luminosity $L_X$ are taken from the selected
models
of Table 4 ($\vartheta_{SN} = 1/4$).

\end{document}